# A monogamy equalities of complementarity relation and distribute entanglement for multi-qubit pure states

**Xin-wei Zha[1],\*, Yanpeng Zhang[2], †**

[1]School of Science, Xi'an University of Posts and Telecommunications, Xi'an, 710121, China

[2]Key Laboratory for Physical Electronics and Devices of the Ministry of Education & Shaanxi Key Lab of Information Photonic Technique, Xi'an Jiaotong University, Xi'an 710049, China

Corresponding author: *zhxw@xupt.edu.cn; †ypzhang@mail.xjtu.edu.cn

**Abstract:** We propose a method to detect genuine quantum correlation for multi-qubit pure states. We then derive a complementarity relations for pure quantum states of N qubits. We prove that in all many-qubit systems there exist strict monogamy laws for quantum correlations. On the other hand, it is known that the entanglement monogamy equality proposed by Coffman, Kundu, and Wootters is in general not true for multiqubit states. Inducing from the CKW equality, we find a proper form of entanglement monogamy equality for arbitrary quantum states. The total quantum correlation of qubit $k$ with the remaining qubits $R_k$ can be characterizes. Furthermore, the quantum correlation of qubit $mn$ with the remaining qubits $R_{mn}$ can also be obtained. Furthermore, some monogamy relations have been obtained.

**1. Introduction**: Quantification of multi-qubit entanglement has remained an outstanding but important challenge in quantum information science. Nowadays, it is a crucial physical resource widely used in quantum-information processing, as in quantum communication [1–2] and quantum computation [3–4].

Complementarity is perhaps the most important phenomenon distinguishing systems that are inherently quantum mechanical from those that may accurately be treated classically. The study of complementary in composite systems has a fairly short history by comparison. Besides the most well-known complementarity principle introduced by Bohr [5], many other kinds of complementarity relation have also been discussed [6-8]. numerous experimental works have been done in this field [9–11].

In particular for two-state systems, elegant relations between two complementary observables have been derived [12-14]. Jakob and Bergou [15] have derived a complementarity relation for an arbitrary pure state of two qubits. Jaeger, et al., [16] have derived a complementarity relation between multipartite entanglement and mixedness for specific classes of N-qubit systems. Cai et al. [17] also established a complementarity relation between local and nonlocal information for two- and three-qubit pure states . For arbitrary n-qubit pure states, they gave following conjecture $\sum_i I_i + 2\sum_{i_1<i_2} \tau_{i_1 i_2} + \cdots + n \sum_{i_1<i_2\cdots<i_n} \tau_{i_1 i_2 \cdots i_n} = n$.

On the other hand, the study of distributed entanglement for multipartite states is important for quantum information In 2000 Coffman, Kundu, and Wootters (CKW) formalized the monogamy of entanglement for a three-qubit system [18] i.e. $\tau_{A(BC)} = \tau_{AB} + \tau_{AC} + \tau_{ABC}$. Despite the importance of monogamy relations for quantum information, the knowledge of quantitative relations for other forms of entanglement is so far rather limited.

In this paper, we propose a method for detecting genuine multipartite correlations for arbitrary *n*-qubit quantum system. We derive a complementarity relations for many-qubit systems by using linear entropy. A monogamy equality analogous to Coffman-Kundu-Wootters (CKW) equality is established.

## 2. Complementarity relations for many-qubit systems

It is known that information contained in multi-qubit systems can be distributed in local and non-local form. According to Brukner-Zeilinger principle: the total information of one qubit is one bit and the total information of N-qubit system is N bit (for pure states) . Therefore, the total information content of a N-qubit system

$$I_{total} = I_{local} + I_{non-local} = N \tag{1a}$$

For an *N*-qubit quantum system in pure state $|\psi\rangle$, the amount of information in local form

$$I_{local}(|\psi\rangle) = \sum_{i=1}^{n} I_i \tag{2}$$

Where $I_i = F_i, i = 1, 2, \cdots, n$ , and that [23-24]

$$F_i = \langle\psi|\sigma_{ix}|\psi\rangle^2 + \langle\psi|\sigma_{iy}|\psi\rangle^2 + \langle\psi|\sigma_{iz}|\psi\rangle^2 \tag{3}$$

We can show that [17] $I_i = 2tr\rho_i^2 - 1$, where $\rho_i = tr\rho_{1\cdots i-1,i+1\cdots n}(|\psi\rangle\langle\psi|)$ is the reduced density operator of the *i*th qubit.

The nonlocal information exists not only in 2-qubit , but also 3-qubit. 4-qubit and n-qubit. We define

$$I_{nonlocal} = (I_{12} + I_{13} + \cdots I_{n-1,n}) + (I_{123} + I_{124} + \cdots I_{n-2,n-1,n}) + \cdots + I_{123\cdots n} \tag{4}$$

Where $I_{ij} = F_{ij} - 1, ij = 12, 13, \cdots n-1, n$ ;

$I_{ijk} = F_{ijk} - 1, ijk = 123, 124, \cdots, n-2, n-1, n$ ; $\cdots$; $I_{123\cdots n} = F_{123\cdots n} - 1$ .

and

$$\begin{aligned}
F_{ij} = & \langle\psi|\sigma_{ix}\sigma_{jx}|\psi\rangle^2 + \langle\psi|\sigma_{ix}\sigma_{jy}|\psi\rangle^2 + \langle\psi|\sigma_{ix}\sigma_{jz}|\psi\rangle^2 \\
& + \langle\psi|\sigma_{iy}\sigma_{jx}|\psi\rangle^2 + \langle\psi|\sigma_{iy}\sigma_{jy}|\psi\rangle^2 + \langle\psi|\sigma_{iy}\sigma_{jz}|\psi\rangle^2 \\
& + \langle\psi|\sigma_{iz}\sigma_{jx}|\psi\rangle^2 + \langle\psi|\sigma_{iz}\sigma_{jy}|\psi\rangle^2 + \langle\psi|\sigma_{iz}\sigma_{jz}|\psi\rangle^2
\end{aligned} \quad (5)$$

$$\begin{aligned}
F_{ijk} = & \langle\psi|\sigma_{ix}\sigma_{jx}\sigma_{kx}|\psi\rangle^2 + \langle\psi|\sigma_{ix}\sigma_{jx}\sigma_{ky}|\psi\rangle^2 + \langle\psi|\sigma_{ix}\sigma_{jx}\sigma_{kz}|\psi\rangle^2 \\
& + \langle\psi|\sigma_{ix}\sigma_{jy}\sigma_{kx}|\psi\rangle^2 + \langle\psi|\sigma_{ix}\sigma_{jy}\sigma_{ky}|\psi\rangle^2 + \langle\psi|\sigma_{ix}\sigma_{jy}\sigma_{kz}|\psi\rangle^2 \\
& + \cdots \\
& + \langle\psi|\sigma_{iz}\sigma_{jz}\sigma_{kx}|\psi\rangle^2 + \langle\psi|\sigma_{iz}\sigma_{jz}\sigma_{ky}|\psi\rangle^2 + \langle\psi|\sigma_{iz}\sigma_{jz}\sigma_{kz}|\psi\rangle^2
\end{aligned}$$

Therefore, the complementarity relation is as follows

$$(I_1 + I_2 + \cdots I_n) + (I_{12} + I_{13} + \cdots I_{n-1,n}) + (I_{123} + I_{124} + \cdots I_{n-2,n-1,n}) + \cdots + I_{123\cdots n} = N \quad (1b)$$

One can easily to show that $I_{local} = N$, $I_{non-local} = 0$ for product state; $I_{non-local} \neq 0$, $I_{local} < N$ for entangled state.

We start by considering the simplest case of a two-qubit system in the pure state

$$|\varphi\rangle_{12} = a_0|00\rangle + a_1|01\rangle + a_2|10\rangle + a_3|11\rangle. \quad (6)$$

From Eq. (1b), we have

$$I_1 + I_2 + I_{12} = 2 \quad (7)$$

We can show that $I_{12} = 2\tau_{12}$, where $\tau_{12}$ is the square of concurrence. Therefore, Eq (7) is just that of Eq (9) of reference[17]. For product state, we have $I_1 = I_2 = 1$, $I_{12} = 0$; For Bell state, that is $I_1 = I_2 = 0$, $I_{12} = 2$.

For three-qubit pure states,

$$|\varphi\rangle_{123} = a_0|000\rangle + a_1|001\rangle + a_2|010\rangle + a_3|011\rangle + a_4|100\rangle + a_5|101\rangle + a_6|110\rangle + a_7|111\rangle \quad (8)$$

From Eq. (1b), we know,

$$I_1 + I_2 + I_3 + I_{12} + I_{13} + I_{23} + I_{123} = 3 \quad (9)$$

For the GHZ state, $|\varphi\rangle_{123} = \frac{1}{\sqrt{2}}(|000\rangle + |111\rangle)$, it is easy to show that $I_1 = I_2 = I_3 = 0$, $I_{12} = I_{13} = I_{23} = 0$, $I_{123} = 3$. For the W state, $|\varphi\rangle_{123} = \frac{1}{\sqrt{3}}(|001\rangle + |010\rangle + |100\rangle)$, we can obtain

$$I_1 = I_2 = I_3 = \frac{1}{9}, \ I_{12} = I_{13} = I_{23} = 0, \ I_{123} = \frac{24}{9}.$$

For four-qubit pure states,

$$|\varphi\rangle_{1234} = a_0|0000\rangle + a_1|0001\rangle + a_2|0010\rangle + a_3|0011\rangle$$
$$+ a_4|0100\rangle + a_5|0101\rangle + a_6|0110\rangle + a_7|0111\rangle$$
$$+ a_8|1000\rangle + a_9|1001\rangle + a_{10}|1010\rangle + a_{11}|1011\rangle \quad (10)$$
$$+ a_{12}|1100\rangle + a_{13}|1101\rangle + a_{14}|1110\rangle + a_{15}|1111\rangle$$

From Eq. (1-4), we know,

$$I_1+I_2+I_3+I_4+(I_{12}+I_{13}+I_{14}+I_{23}+I_{24}+I_{34})+(I_{123}+I_{124}+I_{134}+I_{234})+I_{1234}=4 \quad (11)$$

For the GHZ state, $|\varphi\rangle_{1234} = \frac{1}{\sqrt{2}}(|0000\rangle + |1111\rangle)$, one can have

$I_1 = I_2 = I_3 = I_4 = 0$, $I_{12} = I_{13} = I_{14} = I_{23} = I_{24} = I_{34} = 0$, $I_{123} = I_{124} = I_{134} = I_{234} = -1$, $I_{1234} = 8$.

For the W state, $|\varphi\rangle_{1234} = \frac{1}{2}(|0001\rangle + |0010\rangle + |0100\rangle + |1000\rangle)$, we have

$I_1 = I_2 = I_3 = I_4 = \frac{1}{4}$, $I_{12} = I_{13} = I_{14} = I_{23} = I_{24} = I_{34} = -\frac{1}{2}$, $I_{123} = I_{124} = I_{134} = I_{234} = \frac{3}{4}$, $I_{1234} = 3$.

On the other hand, we can have

$$(I_{12} + I_{13} + I_{14} + I_{23} + I_{24} + I_{34}) - (I_1+I_2+I_3+I_4) = 4(\tau_{1234} - 1) \quad (12)$$

Where $\tau_{1234}$ is n-tangle, which can be defined as [19],

$$\tau_{1234} = |\langle\psi|\sigma_{1y} \otimes \sigma_{2y} \otimes \sigma_{3y} \otimes \sigma_{4y}|\psi^*\rangle|^2$$

## 3  A monogamy equality analogous to Coffman-Kundu-Wootters (CKW) equality

It is well know that the total quantum correlation of qubit $k$ with the remaining qubits $R_k$ can be characterizes by the linear entropy

$$\tau_{k(R_k)} = 2(1 - tr(\rho_k^2)). \quad (13)$$

For a two-qubit pure state, the linear entropy is a bipartite quantum correlation. For a three-qubit case, the $\tau_{k(R_k)}$ is composed of the two-qubit and genuine three-qubit correlations [19]. For an N-qubit pure state, the linear entropy is contributed by the different levels of quantum correlation [20].

For an N-qubit pure state, we presented a natural generalization that the linear entropy is contributed by different levels of quantum correlations, i.e.,

$$(2^{n-2}+1)\tau_{1(R_1)} = I_{12}+I_{13}\cdots I_{123}+I_{124}+\cdots+I_{123\cdots n} \tag{14}$$

For a two-qubit pure state, the linear entropy is

$$2\tau_{1(2)} = I_{12} \tag{15}$$

For a three-qubit case, the linear entropy is

$$3\tau_{1(23)} = I_{12}+I_{13}+I_{123} \tag{16}$$

For a four-qubit case, the linear entropy is

$$5\tau_{1(234)} = I_{12}+I_{13}+I_{14}+I_{123}+I_{124}+I_{134}+I_{1234} \tag{17}$$

For a five-qubit case, the linear entropy is

$$9\tau_{1(2345)} = (I_{12}+I_{13}+I_{14}+I_{15})+(I_{123}+I_{124}+I_{125}+I_{134}+I_{135}+I_{145})$$
$$+(I_{1234}+I_{1235}+I_{1245}+I_{1345})+I_{12345} \tag{18}$$

Furthermore, we define that the total quantum correlation of qubit $ml$ with the remaining qubits $R_{ml}$ can be characterizes by the linear entropy

$$\tau_{ml(R_{ml})} = 2(1-tr(\rho_{ml}^2)). \tag{19}$$

Then, for an N-qubit pure state, we can obtain that the linear entropy is contributed by different levels of quantum correlations, i.e.,

$$2(2^{n-4}+1)\tau_{12(34\cdots n)} = I_{13}+I_{14}\cdots+I_{1n}+I_{23}+I_{24}\cdots+I_{2n}+I_{123}+I_{124}+\cdots+I_{12n}\cdots+I_{123\cdots n}, n\geq 4 \tag{20}$$

For four-qubit pure states, it can be expressed

$$4\tau_{12(34)} = (I_{13}+I_{14}+I_{23}+I_{24})+(I_{123}+I_{124}+I_{134}+I_{234})+I_{1234} \tag{21}$$

Similarly, for five-qubit pure states, we can have

$$6\tau_{12(345)} = (I_{13}+I_{14}+I_{15}+I_{23}+I_{24}+I_{25})+(I_{123}+I_{124}+I_{125}+I_{134}+I_{135}+I_{145}+I_{234}+I_{235}+I_{245})$$
$$+(I_{1234}+I_{1235}+I_{1245}+I_{1345}+I_{2345})+I_{12345} \tag{22}$$

## 4. Discussions and conclusions

If the system is in a mixed state, the above equation will be replaced by an inequality, i.e.

$$I_{total} = I_{local} + I_{non-local} \leq N \tag{23}$$

In ref[21], they give relation $tr\rho_1^2 + tr\rho_2^2 - tr\rho_{12}^2 \leq 1$.

Recently, Christopher Eltschka and Jens Siewert [22] have given some degree-2 and degree-4 monogamy relations. As reference [22], using this method, we can get

$$tr\rho_1^2 + tr\rho_2^2 - tr\rho_{12}^2 = 1 - tr(\rho_{12}\tilde{\rho}_{12}) \quad (24)$$

Here $\tilde{\rho}_{12} = (\sigma_y \otimes \sigma_y)\rho_{12}^*(\sigma_y \otimes \sigma_y)$ is the time-reversed density matrix of $\rho_{12}$.
Because of $tr(\rho_{12}\tilde{\rho}_{12}) \leq 1$ .it is obvious that $tr\rho_1^2 + tr\rho_2^2 - tr\rho_{12}^2 \leq 1$.
Similarly, we can have

$$tr\rho_{123}^2 - \frac{1}{2}\left(tr\rho_{12}^2 + tr\rho_{13}^2 + tr\rho_{23}^2 + tr(\rho_{12}\tilde{\rho}_{12}) + tr(\rho_{13}\tilde{\rho}_{13}) + tr(\rho_{23}\tilde{\rho}_{23})\right) + \frac{3}{2} = 1 - tr(\rho_{123}\tilde{\rho}_{123}) \quad (25)$$

i.e. $tr\rho_{123}^2 - \frac{1}{2}\left(tr\rho_{12}^2 + tr\rho_{13}^2 + tr\rho_{23}^2 + tr(\rho_{12}\tilde{\rho}_{12}) + tr(\rho_{13}\tilde{\rho}_{13}) + tr(\rho_{23}\tilde{\rho}_{23})\right) + \frac{3}{2} \geq 0$

On the other hand, for a four-qubit, using Eq（16）and（20）,one can obtain

$$\left(5\tau_{1(234)} + 5\tau_{2(134)} + 5\tau_{3(124)} + 5\tau_{4(123)}\right) - \left(4\tau_{12(34)} + 4\tau_{13(24)} + 4\tau_{14(23)}\right) = I_{1234} \quad (26)$$

In conclusion, we found a complementarity relations for pure quantum states of N qubits. We prove that in all many-qubit systems there exist strict monogamy laws for quantum correlations. We may also find a monogamy equality analogous to Coffman-Kundu-Wootters (CKW) equality. Furthermore, some monogamy equality is established. We believe those result can play an important role in quantum communication and computing.

Acknowledgments

This work is supported by the Natural Science Foundation of Shaanxi Province of China (Grant No. 2013JM1009).